\documentclass[fleqn,usenatbib]{mnras}
\usepackage[T1]{fontenc}
\usepackage{ae,aecompl}

\usepackage{graphicx}   % Including figure files
\usepackage{amsmath}    % Advanced maths commands
\usepackage{amssymb}    % Extra maths symbols
\usepackage[utf8]{inputenc}
\usepackage[varg]{txfonts}
\usepackage{amsfonts}
\usepackage{nicefrac}
\usepackage{float}
\usepackage{graphicx}
\usepackage{textcomp}
\usepackage{pdfpages}
\usepackage{url}
\usepackage{natbib}
\usepackage{hyperref}
\usepackage{prettyref}
\newrefformat{fig}{\text{Fig.~\ref{#1}}}
\newrefformat{eq}{\text{Eq.~\ref{#1}}}
\newrefformat{tab}{\text{Table~\ref{#1}}}
\newrefformat{sec}{\text{Sect.~\ref{#1}}}
\newrefformat{enu}{\text{~\ref{#1}}}

\title[Porosity: Influence on debris disc analysis]{The influence of dust grain porosity on the analysis of debris disc observations}

\author[R. Brunngräber et al.]{
Robert Brunngräber\thanks{rbrunngraeber@astrophysik.uni-kiel.de}$^{1}$, Sebastian Wolf$^{1}$, Florian Kirchschlager$^{1}$ and Steve Ertel$^{2,3}$
\\
$^{1}$Institut für theoretische Physik und Astrophysik, Christian-Albrechts-Universität zu Kiel, Leibnizstr. 15, 24118 Kiel, Germany\\
$^{2}$European Southern Observatory, Alonso de Cordova 3107, Vitacura, 19001 Casilla, Santiago 19, Chile\\
$^{3}$Steward Observatory, Department of Astronomy, University of Arizona, 933 North Cherry Avenue, Tucson, AZ 85721, USA
}

\date{Accepted XXX. Received YYY; in original form ZZZ}

\pubyear{2016}

\begin{document}
\label{firstpage}
\pagerange{\pageref{firstpage}--\pageref{lastpage}}
\maketitle

\begin{abstract}
Debris discs are often modelled assuming compact dust grains, but more and more evidence for the presence of porous grains is found.
We aim at quantifying the systematic errors introduced when modelling debris discs composed of porous dust with a disc model assuming spherical, compact grains.
We calculate the optical dust properties derived via the fast, but simple effective medium theory. The theoretical lower boundary of the size distribution -- the so-called 'blowout size' -- is compared in the cases of compact and porous grains. Finally, we simulate observations of hypothetical debris discs with different porosities and feed them into a fitting procedure using only compact grains. The deviations of the results for compact grains from the original model based on porous grains are analysed.
We find that the blowout size increases with increasing grain porosity up to a factor of two. An analytical approximation function for the blowout size as a function of porosity and stellar luminosity is derived. The analysis of the geometrical disc set-up, when constrained by radial profiles, are barely affected by the porosity. However, the determined minimum grain size and the slope of the grain size distribution derived using compact grains are significantly overestimated. Thus, the unexpectedly high ratio of minimum grain size to blowout size found by previous studies using compact grains can be partially described by dust grain porosity, although the effect is not strong enough to completely explain the trend.
\end{abstract}


%%%%%%%%%%%%%%%%%%%%%%%%%%%%%%%%%%%%%%%%%%%%%%%%%%

%%%%%%%%%%%%%%%%% BODY OF PAPER %%%%%%%%%%%%%%%%%%

\section{Introduction}
\label{sec:intro}
    In the analysis of debris disc observations dust grains are often considered as homogeneous, compact spheres \citep{ertel-et-al-2012,ertel-et-al-2014,schueppler-et-al-2014,marshall-et-al-2014,pawellek-et-al-2014,kral-et-al-2015}, although it is expected that dust particles in circumstellar discs are neither compact, homogeneous nor spherical but possess complex, irregular shapes and inclusions of various materials and voids \citep{dominik-tielens-1997,blum-et-al-2000,krugel-2003,ormel-et-al-2008}. In addition, recent studies with high-contrast, high-resolution polarimetric imaging with \textit{VLT}/NaCo, \textit{VLT}/SPHERE and \textit{Gemini}/GPI also show that the measured data cannot be reproduced by compact, spherical dust grains \citep{milli-et-al-2015}, because these grain irregularities have various implications on the optical properties of the dust, such as on its absorption and scattering cross sections, and its ability to emit and scatter polarized light, and, hence, on the resulting appearance of a debris disc \citep{min-et-al-2012}. Besides the optical properties of the dust, the appearance depends on its spatial distribution. Usually, only very weak constraints on the dust grain properties can be derived directly, i.e. independently from the underlying disc model. Hence, the dust model is mostly chosen to be as simple as possible. Besides, this approach is much less computationally expensive than modelling under the assumption of more complex shapes and/or compositions. However, the influence of these assumptions on the model results needs to be investigated.
    
    In many previous studies, authors studied the influence of porous, or fluffy, grains on the interstellar extinction, absorption and scattering efficiencies, dust temperature, opacity, or the shape and position of infrared bands. It was found that these properties can in principle be used to distinguish between compact and porous grains in protoplanetary discs as well as in debris discs \citep{voshchinnikov-et-al-2006,voshchinnikov-et-al-2007,min-et-al-2008,kataoka-et-al-2014,kirchschlager-wolf-2013,kirchschlager-wolf-2014}. However, only a few debris disc modelling studies exist to date in which porous dust is used \citep{augereau-et-al-1999a,augereau-et-al-1999b,li-lunine-2003,churcher-et-al-2011,acke-et-al-2012,lebreton-et-al-2012,donaldson-et-al-2013,seok-li-2015}. Although the results are promising, the effect of porosity on the derived disc parameters has not been investigated yet.
    
    Moreover, porosity influences the physical behaviour, and hence the orbit of the particles as well. In an optically thin disc, the radiation pressure force can be large enough to expel very small grains \citep{burns-et-al-1979}. The critical grain size is called 'blowout size' and is expected to be the lower limit of the grain size distribution. However, many attempts to fit observations of debris discs suggest a minimum grain size larger than the expected blowout size by a factor of 5 -- 10 \citep{roccatagliata-et-al-2009,ertel-et-al-2011,loehne-et-al-2012}, whereas collision theory is able to explain only a factor of 2 -- 3 \citep{krivov-et-al-2006,thebault-augereau-2007,thebault-wu-2008}. In \citet{pawellek-et-al-2014}, the authors analysed 34 \textit{Herschel}-resolved debris discs and found that the minimum grain size in their best-fitting models are up to ten times larger than expected for stars with a luminosity comparable to the sun. They also found that the deviation of the minimum grain size from the blowout limit decreases for increasing luminosity. In a subsequent study, \citet{pawellek-krivov-2015} found that this trend is robust against different grain compositions and a simple porosity model.
    
    In this study, we quantify the influence of porous dust grains on the analysis of observations in the case of optically thin debris discs. In \prettyref{sec:dust_model}, our porous dust model is described. \prettyref{sec:blowout} shows the blowout size as a function of grain porosity and stellar luminosity, as well as the deviations from commonly used approximations. The last and major part of this article concerns the systematic error introduced during disc modelling under the assumption of compact dust grains in \prettyref{sec:fit}. Thus, we show whether porosity introduces differences in the resulting disc parameter values and how strong these differences are.
    
\section{Porosity and optical dust properties}
\label{sec:dust_model}
    \subsection{Porosity model}
    \label{sec:por}
        In this study, we assume that the dust grains have a spherical shape with radius $s$ and are purely composed of astronomical silicate (astrosil), with a bulk density of $\varrho_{\text{0}} = 3.5$~g~cm$^{-3}$ \citep{draine-2003a}. The optical data are taken from \citet{draine-2003b,draine-2003c}. With $V_{\text{dust}}$, $V_{\text{vacuum}}$ and $V_{\text{total}}$ denoting the volumes taken up by the dust grains and voids and the total volume of the sphere, respectively, the porosity of these grains is defined as follows:
        \begin{equation}
            \mathcal{P} = 1 - \frac{V_{\text{dust}}}{V_{\text{total}}} = \frac{V_{\text{vacuum}}}{V_{\text{total}}}\ ,
        \end{equation}
        where $\mathcal{P} = 0$ corresponds to a compact, spherical grain composed of astrosil and $\mathcal{P} = 1$ to pure vacuum. The mass of a particle with radius $s$ is thus
        \begin{equation*}
            m = \frac{4}{3}\,\pi s^3\ \varrho_{\text{0}}\,\left(1-\mathcal{P}\right)\ .
        \end{equation*}
        We consider porosities between 0.0 and 0.9 in steps of 0.1 to investigate the influence on the observable appearance of debris discs.
    
    \subsection{Optical properties of porous grains}
    \label{sec:optical_prop}
        The well-known Mie theory \citep{mie-1908} is not valid for grains with irregular shapes and inhomogeneities or inclusions. Thus, approximations have to be used to calculate optical properties of porous dust grains. The resulting properties and hence all further analysis may depend on the used approximation method. A comparison between the two most commonly used methods, the simple but fast effective mixing theory (EMT), and the more sophisticated discrete dipole approximation (DDA;  \citealt{purcell-pennypacker-1973,draine-1988}) has been carried out by a large number of authors \citep{ossenkopf-1991,voshchinnikov-et-al-2005,kirchschlager-wolf-2013}. The optical properties obtained with the EMT are in good agreement with that derived by DDA, and DDA is computationally very expensive, and thus not applicable for the huge parameter space we were considering (see \prettyref{tab:simu_disc} in \prettyref{sec:debris}). Therefore, the computation of the absorption cross sections $C_{\text{abs}}$, scattering cross sections $C_{\text{sca}}$, and the asymmetry factors $g$ in this study are done using the EMT. The refractive indices $n$ and $k$ of astrosil and vacuum are mixed using the Bruggeman mixing rule \citep{bruggeman-1935} to form an 'effective'  material that has the refractive indices of dust with porosity $\mathcal{P}$:
        \begin{equation}
            (1-\mathcal{P})\frac{\varepsilon_{\text{astrosil}}-\varepsilon_{\text{eff}}}{\varepsilon_{\text{astrosil}}+2\varepsilon_{\text{eff}}} + \mathcal{P}\frac{\varepsilon_{\text{vacuum}}-\varepsilon_{\text{eff}}}{\varepsilon_{\text{vacuum}}+2\varepsilon_{\text{eff}}} = 0
        \end{equation}
        with the complex permittivity $\varepsilon = (n^2 - k^2) + i\cdot(2nk)$, while $\varepsilon_{\text{astrosil}}$, $\varepsilon_{\text{vacuum}}$ and $\varepsilon_{\text{eff}}$ denote the permittivity of the dust, voids, and the resulting effective medium, respectively. In addition, the refractive indices $n$ and $k$ of vacuum are equal to one and zero, respectively, hence $\varepsilon_{\text{vacuum}} \equiv 1$. Afterwards, the optical cross sections and the asymmetry factor $g$ are calculated by the tool \textsc{miex} \citep{wolf-voshchinnikov-2004} using standard Mie theory.
    
\section{Blowout limit}
\label{sec:blowout}
    As mentioned in \prettyref{sec:intro}, the minimum grain size found in debris discs is often larger than the value that is expected from theoretical studies, pointing either to poorly understood destruction mechanisms or to an incorrect treatment of the radiation pressure. In this section, we calculate the influence of porosity on the radiation pressure force on the particles and hence on the minimum grain size that can be expected in the disc. Furthermore, we compare our results to a widely used approximation equation (\prettyref{eq:s_blow_burns}) for the blowout size.
    
    Due to radiation pressure, the orbit of small particles differs from those affected only by gravitation \citep{burns-et-al-1979}. Both, radiation pressure force $F_{\text{rp}}$ and gravitational pull $F_{\text{grav}}$ are in direct proportion to $r^{-2}$. Thus, the ratio is independent of the distance to the star and depends on the properties of the star and the dust only. The so-called $\beta$-ratio is thus given by:
    \begin{equation}
    \label{eq:beta}
        \beta := \frac{F_{\text{rp}}}{F_{\text{grav}}} = \frac{C_{\star}}{s^3\varrho_{\text{0}}\left(1-\mathcal{P}\right)}~\int_{0}^{\infty}{C_{\text{rp}}(s,\mathcal{P},\lambda)\cdot B_{\lambda}\left(T_{\text{eff}}\right)~\text{d}\lambda}\ ,
    \end{equation}
    with
    \begin{equation*}
        C_{\star} = \frac{3}{4\,c}\frac{R_{\star}^2}{GM_{\star}}\ ,
    \end{equation*}
    where $C_{\text{rp}} = C_{\text{abs}} + C_{\text{sca}}\left(1-g\right)$ is the radiation pressure cross section, $c$ the speed of light, $G$ the gravitational constant, $B_{\lambda}$ the Planck function, and $R_{\star}$, $M_{\star}$, $T_{\text{eff}}$ are the stellar radius, mass, and effective temperature, respectively \citep{burns-et-al-1979,koehler-mann-2002,kirchschlager-wolf-2013}. Contrary to the common notation in the literature, we denote the radiation pressure cross section with the index 'rp' instead of 'pr', as the latter can be easily misinterpreted as a cross section related to the Poynting--Robertson drag \citep{poynting-1904,robertson-1937,wyatt-whipple-1950}. To calculate the stellar mass we use the common relation for main-sequence stars $M_{\star}\propto L_{\star}^{\nicefrac{1}{3.8}}$ \citep{eddington-1924,unsoeld-baschek-2013}.
    
    The equation of motion, taking the Newtonian law of gravity and the radiation pressure into account, is hence
    \begin{equation}
    \label{eq:equ_motion}
        m\ddot{r} = -G\frac{M_{\star}m\left(1-\beta\right)}{r^2}\ .
    \end{equation}
    \prettyref{eq:equ_motion} results in unbound parabolic/hyperbolic orbits for dust grains with $\beta \ge 1$. The smallest grains in debris discs are produced either by collision of two larger grains or by ejection from a large parent body, e.g. a comet. Due to the higher velocities of their parent bodies, the newly born particles can have unbound orbits although they have a $\beta$-value less than one. The value of this blowout limit has been derived by \citet{burns-et-al-1979} to be $\beta_{\text{blow}} = 0.5$ assuming circular orbits of the parent body. The corresponding grain size is called blowout size $s_{\text{blow}}$ and determines, in first-order approximation, the smallest grain present in the debris disc.
    
\begin{figure}
        \centering
        \resizebox{\hsize}{!}{\includegraphics{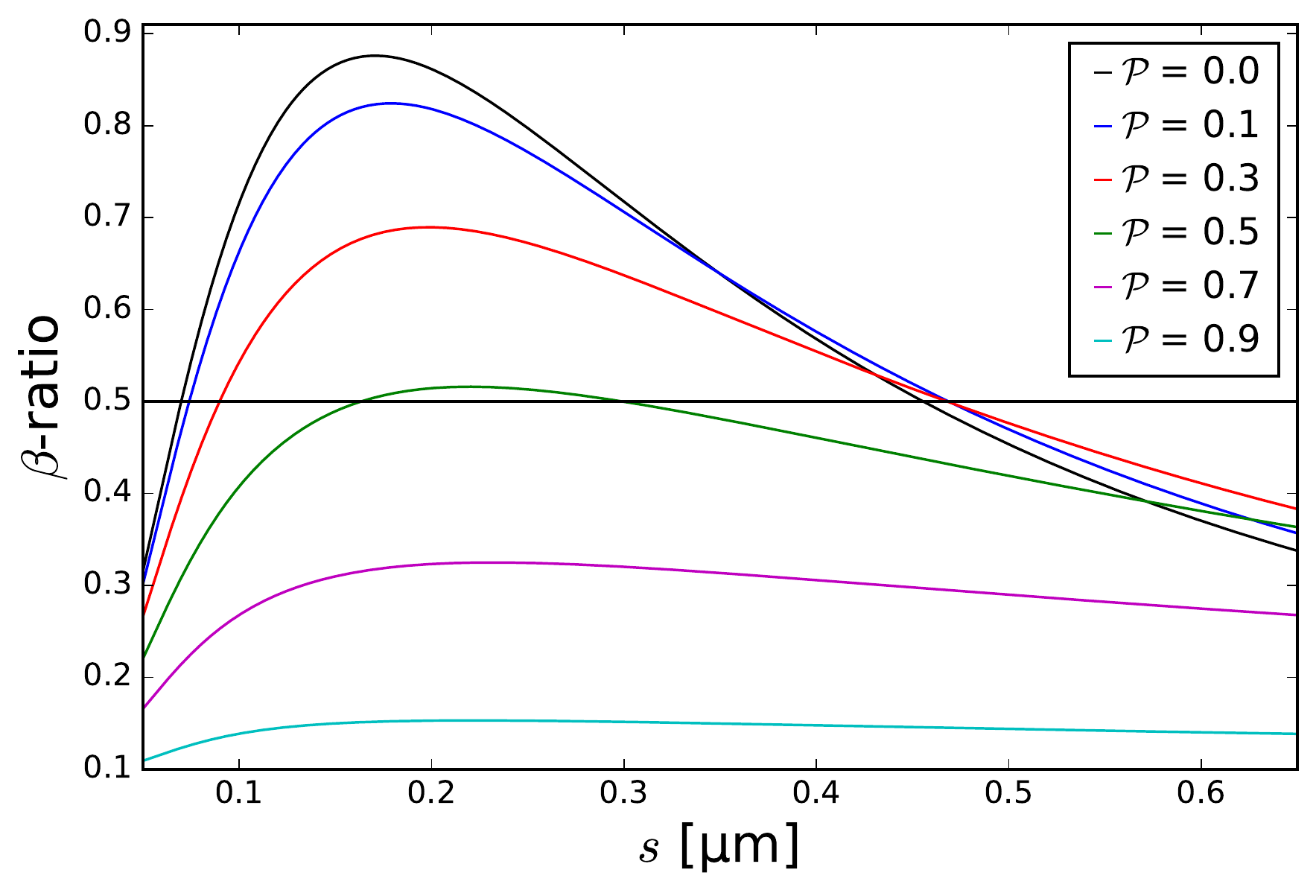}}\\
        \resizebox{\hsize}{!}{\includegraphics{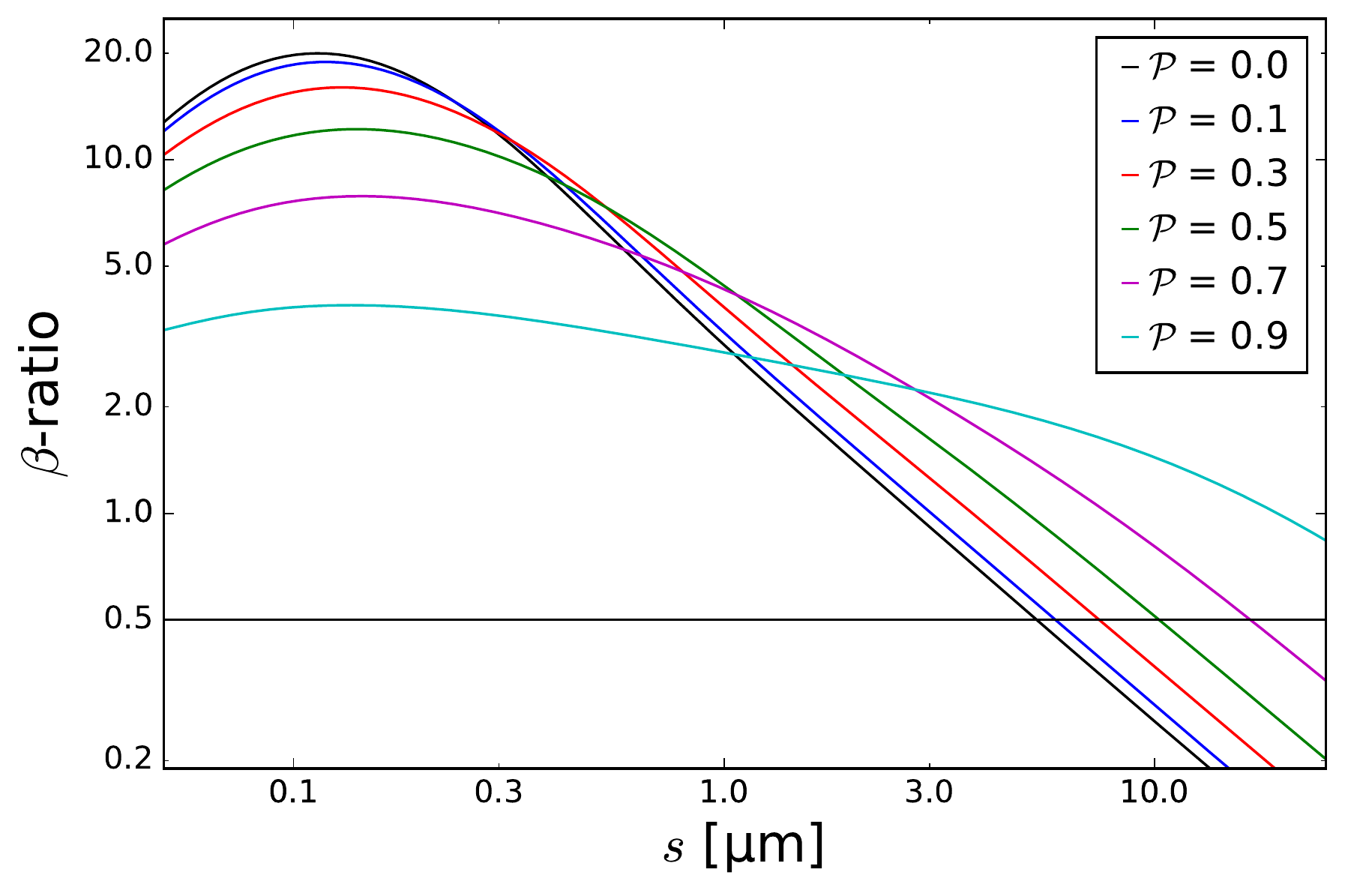}}\\
        \caption[caption]{Ratio of radiation pressure and gravitation $\beta$ for two different spectral types.\\\hspace{\textwidth}
        \textit{Top:} Solar-type star with $T_{\text{eff}} = 5770$~K and $L_{\star} = 1~\text{L}_{\sun}$;
        \textit{Bottom:} Fomalhaut-like star with $T_{\text{eff}} = 8195$~K and $L_{\star} = 15.5~\text{L}_{\sun}$. Different colours indicate different grain porosities $\mathcal{P}$. Indicated by the black, horizontal line is the blowout limit $\beta_{\text{blow}} = 0.5$.}
    \label{fig:beta}
    \end{figure}

    In \prettyref{fig:beta}, the ratio of gravitational pull and radiation pressure $\beta$ is plotted for two different stellar types, a solar-type star (spectral type G2V) in the upper panel and a Fomalhaut-like star (spectral type A3V) in the lower panel. For larger grains, $\beta$ is higher if the porosity is higher. This is due to the factor $(1-\mathcal{P})^{-1}$ in \prettyref{eq:beta}. However, for smaller grains, i.e. grains with sizes comparable to the wavelength of maximum stellar emission, the radiation pressure cross section $C_{\text{rp}}$ decreases for larger porosities and thus $\beta$ drops to smaller values. For cooler stars, this results in an increasing blowout size for small porosities and decreasing blowout size for higher porosities. In the case of the solar-type star, the maximum blowout size is $s_{\text{blow}} \approx 0.48$~\textmu m for a porosity of about $\mathcal{P} = 0.2$. If the porosities are larger than $0.5$, the blowout limit does not exist, i.e. $\beta < 0.5$ for all radii. For hotter stars, $s_{\text{blow}}$ is monotonically increasing with increasing porosity.
    
\begin{figure}
        \centering
        \resizebox{\hsize}{!}{\includegraphics{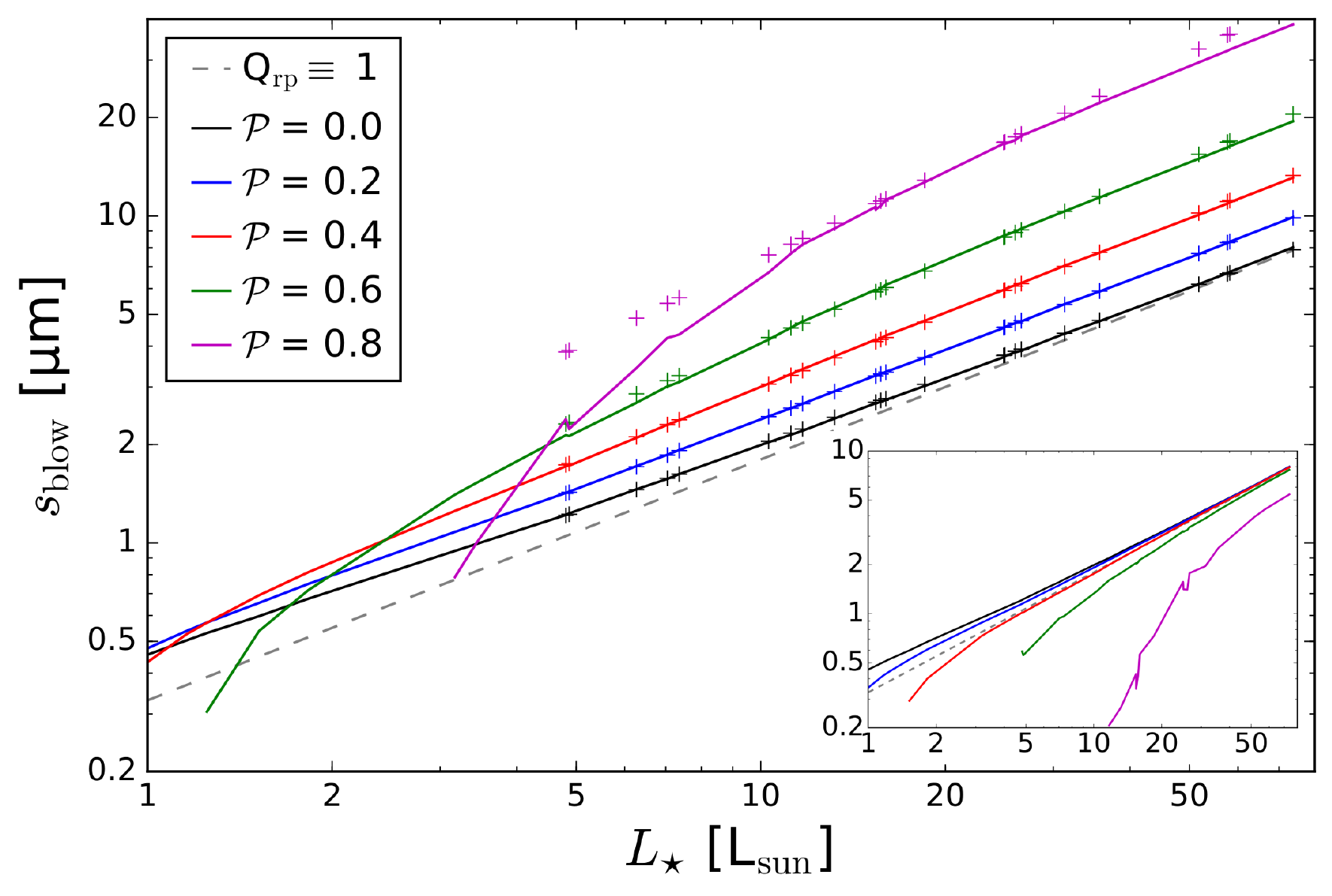}}
        \caption[caption]{Blowout size $s_{\text{blow}}$ calculated via \prettyref{eq:beta} as a function of stellar luminosity. Different colours indicate different grain porosities $\mathcal{P}$. For comparison, the dashed, grey line shows $s_{\text{blow}}$ obtained from the approximation \prettyref{eq:s_blow_burns}. Additionally, crosses mark the blowout sizes that are obtained with the analytical expression \prettyref{eq:sblow}.\\
        \textit{Inset:} The blowout size corrected for the different grain densities, i.e. the factor $(1-\mathcal{P})^{-1}$ in \prettyref{eq:beta} is not considered. See text for further explanation.}
    \label{fig:sblow_l}
    \end{figure}

        We compare $s_{\text{blow}}$ obtained from \prettyref{eq:beta} with the commonly used approximation \citep{burns-et-al-1979,artymowicz-1988,deller-maddison-2005,hahn-2010,pawellek-et-al-2014,veras-et-al-2015}
    \begin{equation}
    \label{eq:s_blow_burns}
        \frac{s_{\text{blow}}}{1\text{~\textmu m}} = 0.33\,\left(\frac{3.5\,\text{g cm}^{-3}}{\varrho}\right)\left(\frac{L_{\star}}{\text{L}_{\sun}}\right)\left(\frac{M_{\star}}{\text{M}_{\sun}}\right)^{-1}\ ,
    \end{equation}
    where the radiation pressure coefficient $Q_{\text{rp}}=\nicefrac{C_{\text{rp}}}{\pi s^2}\equiv1$ for all wavelengths and grain sizes, rejecting all information about the wavelength dependence of the optical dust properties. To calculate the blowout size, we use the stellar properties of main-sequence stars with luminosities ranging from 0.012~L$_{\sun}$ to 73.8~L$_{\sun}$, which are taken from table 2 in \citet{pawellek-et-al-2014}. In \prettyref{fig:sblow_l}, the blowout size as a function of stellar luminosity $L_{\star}$ is shown. The blowout sizes obtained from \prettyref{eq:s_blow_burns} (in \prettyref{fig:sblow_l} refered to as '$Q_{\text{rp}} \equiv 1$') clearly underestimate the blowout size for compact grains for almost all luminosities. For a luminosity of $L_{\star}\approx\text{L}_{\sun}$, it is only 72 per cent of the value obtained with \prettyref{eq:beta}. However, for the brightest stars considered the deviation between blowout size and approximation diminishes to less than 2 per cent. Furthermore, it becomes clear that the slope of the approximation is systematically too high if compared to the more strict definition of \prettyref{eq:beta} for compact grains. The inset in \prettyref{fig:sblow_l} shows the blowout size corrected for different grain densities, i.e. the factor $(1-\mathcal{P})^{-1}$ in \prettyref{eq:beta} is not considered. Therefore, the visible deviations for different porosities are due to variations of $C_{\text{rp}}$ only. Especially for low luminosities, the deviations can exceed 200 per cent. For the highest luminosities, the wavelength of maximum stellar emission is much smaller than the grain size, and thus we reach the limit of geometrical optics where $Q_{\text{abs}}$ is close or equal to one for all porosities.
    
    To provide a simple parametrization of the blowout size as a function of stellar luminosity and dust grain porosity, we fit the power law $s_{\text{blow}} = a\cdot L_{\star}^{b}$ with $a$ and $b$ as fit parameters to the results of \prettyref{eq:beta}. Please note, that the fit was done for $L_{\star}\gtrsim5$ L$_{\sun}$ and $\mathcal{P}\le0.5$ only, because for lower luminosities or higher porosities the results do not suggest a simple power law; see \prettyref{fig:sblow_l}. Further, we find that the results for $a$ and $b$ can be written as a function of porosity. Thus, the blowout size $s_{\text{blow}}$ of porous grains can be analytically expressed as
    \begin{equation}
    \label{eq:sblow}
        \frac{s_{\text{blow}}}{1\text{~\textmu m}} = \left(\frac{3.5\,\text{g cm}^{-3}}{\varrho}\right)\,a_1\,(1-\mathcal{P})^{b_1}\cdot \left(\frac{L_{\star}}{\text{L}_{\sun}}\right)^{a_2\,(1-\mathcal{P})^{b_2}}
    \end{equation}
    with
    \begin{align*}
        a_1 &= 0.414  \pm 0.004\ ,\\
        b_1 &= -0.508 \pm 0.025\ ,\\
        a_2 &= 0.685  \pm 0.002\ ,\\
        b_2 &= -0.168 \pm 0.008\ .
    \end{align*}
    The blowout sizes calculated with \prettyref{eq:sblow} are marked with crosses in \prettyref{fig:sblow_l} and are in good agreement with the results calculated with \prettyref{eq:beta} for $L_{\star}\gtrsim5\text{ L}_{\sun}$ and $\mathcal{P}\lesssim0.6$. For porosities larger than $\approx0.6$, this equation does not reproduce the numerical value from \prettyref{eq:beta}. Please note, that the values of $a_i$ and $b_i$ are obtained for astrosil only, and may be different for other dust compositions. Furthermore, the blowout limit $\beta_{\text{blow}} = 0.5$ is only valid for dust grains that were ejected from a parent body on a circular orbit. For eccentric orbits, the blowout limit may be smaller or larger, depending on the true anomaly at the time of particle ejection \citep{burns-et-al-1979}.
    
    In \prettyref{eq:beta}, we use the Planck function as an approximation of the real stellar radiation. We investigate the deviations introduced by this assumption by calculating $s_{\text{blow}}$ for a solar-type ($T_{\text{eff}} = 5800$~K, $\log{g}=4.5$, [Fe/H] $=0.0$) and a Vega-like ($T_{\text{eff}} = 9600$~K, $\log{g}=4.0$, [Fe/H] $=-0.5$) star with synthetic stellar spectra from the \textit{Göttingen Spectral Library} \citep{husser-et-al-2013}. We find that the resulting blowout sizes differ by less than 2 per cent from the results with black bodies for all porosities. Hence, the approximation of stars as perfect emitters is justified. In the case of a solar-type star and a porosity of $\mathcal{P}=0.5$, no blowout size exist for the synthetic spectrum because the $\beta$-value is always lower than 0.5, due to the lack of radiation of the solar spectrum at the wavelengths of maximum black-body emission. For even larger porosities, no blowout size exist even for the black body assumption.
    
\section{Implications on the debris disc analysis}
\label{sec:fit}
    To fit and interpret the physical parameters of a debris disc from observations, one must apply several assumptions about the shape and composition of the dust particles. In the simplest case dust grains are approximated as compact spheres composed of a single material, e.g. astrosil; see \prettyref{sec:intro}. However, if the disc is made of porous instead of compact grains, the derived best-fit parameter values do most likely differ from the real ones. In this section, we show how large the deviations from the true parameter values of porous dust are, when we assume compact grains in the analysis of selected observational data. For this purpose we perform the following procedure:
    \begin{enumerate}
        \item \label{enu:1} Calculating the re-emission SED and images of an analytical dust distribution for different porosities $\mathcal{P}$,
        \item Convolution of the images with a circular Gaussian to mimic a real observation; Extraction of radial profiles; Superposition of artificial uncertainties (noise),
        \item Fitting the SED and profiles with a debris disc fitting software assuming compact dust grains, i.e. $\mathcal{P} \equiv 0.0$,
        \item Comparison of the fit results of the different porosities with the known, true values, which were used in \prettyref{enu:1}.
    \end{enumerate}
    Because of the limited parameter space considered, e.g. dust composition and porosity models, this approach is not suitable to provide comprehensive, qualitative statements. However, the goal is to identify general trends of the impact of porous dust grains on the observables of debris discs and the resulting deviations on their interpretation.
    
    The first two steps as well as the disc set-up are discribed in \prettyref{sec:debris}, the third step in \prettyref{sec:sand}, and the results are discussed in \prettyref{sec:results}.
    
    \subsection{Disc set-up and simulated observations}
    \label{sec:debris}
        The disc has a typical spread from 40~au to 200~au and the volume density distribution follows a power law, $r^{-\alpha}$ with $\alpha=1$. The grain sizes are in the range of 3.7~\textmu m and 1~mm with a grain size distribution of $n(s) \propto s^{-q}$, where $n\cdot$d$s$ is the number of particles in the radius intervall $[s,s+\text{d}s]$. The power law index $q$ is set to $3.5$, which is the theoretical value for a collisionally dominated debris disc \citep{dohnanyi-1969}. Due to this steep size distribution, the appearance of the disc is mainly dominated by the smallest grains, and thus the value of $s_{\text{max}} = 1$~mm has a negligible impact and is an often used upper boundary in debris disc analyses \citep{loehne-et-al-2012,pawellek-et-al-2014,rodigas-et-al-2015}. The minimum grain size was chosen to be in the mid-range of the calculated blowout sizes. This set-up is used for ten different porosities ranging between $\mathcal{P}=0.0$ (compact grains) and $\mathcal{P}=0.9$ in steps of $0.1$. In the second column of \prettyref{tab:simu_disc} the considered disc and dust set-up is listed. The used stellar properties are shown in \prettyref{tab:luminosity}. The system has a distance of 8~pc, which is comparable to the distance of the two famous, debris disc hosting stars Fomalhaut and Vega \citep{perryman-et-al-1997}.
        
        We calculate the SED at six wavelengths, logarithmically spaced from 10~\textmu m to 2~mm. Additionally, images are calculated for wavelengths of 70~\textmu m and 160~\textmu m, the wavelengths where the \textit{Herschel}/PACS instrument was sensitive to \citep{poglitsch-et-al-2010}. These calculations are done with the \textsc{debris} tool \citep{ertel-et-al-2011,ertel-diss-2012}.
        
        Subsequently, the images are convolved with a circular 2-D Gaussian with a FWHM equal to the resolution of a telescope with an aperture of 3.5~m, e.g. the \textit{Herschel Space Observatory} \citep{pilbratt-et-al-2010}. From these convolved images radial brightness profiles are extracted. These profiles and SED are considered as our simulated observations of discs made of porous grains. We also introduce artificial uncertainties, which are set to 10 per cent of the 'measured' data of SED and radial profile.
    
    \subsection{SED and profile fit}
    \label{sec:sand}
        In the second step, we use the tool \textsc{SAnD} \citep{ertel-diss-2012,ertel-et-al-2012,loehne-et-al-2012,ertel-et-al-2014,marshall-et-al-2014}, which makes use of the simulated annealing approach to fit SED and radial profiles, and the chi-squared distribution to evaluate each step of the fit process. Note, that from this point on we always use the optical properties of compact astrosil, i.e. $\mathcal{P} = 0.0$, to fit the simulated observations. Thus, similar to the often applied approach to fit observational data of debris discs, we use a dust model that is wrong but easier to calculate.
        
        The parameter space can be found in the three rightmost columns of \prettyref{tab:simu_disc}. Stellar temperature and luminosity, inclination, porosity and maximum grain size are fixed as shown in the table. As the fluxes are in direct proportion to the dust mass for optically thin discs, the latter is determined by scaling the calculated SED to the observations. The inner and outer radius $R_{\text{in}}$ and $R_{\text{out}}$, respectively, the radial density profile $\alpha$, the minimum grain size $s_{\text{min}}$ as well as the slope of the grain size distribution $q$ are the free fit parameters. The fit procedure stops, if a given chi-squared value is reached and no further improvement could be found after a certain number of runs.
        
        For every porosity and star, we run the fit routine with five different starting points of the Markov chain. The results of these five runs are averaged in terms of the arithmetic mean to minimize the intrinsic, statistical uncertainty of a single fit process.
        
        \begin{table}
            \caption{Debris disc parameters.}
            \label{tab:simu_disc}
            \centering
            \begin{tabular}{llllr}
                \hline\hline
                \rule{0pt}{2.5ex}Disc parameter                 & Sim. observation & \multicolumn{3}{c}{\textsc{SAnD}} \\
                                                                &                  & Min. & Max. & n                   \\[1mm]
                \hline
                \rule{0pt}{2.5ex}$M_{\text{dust}}$~[$\text{M}_{\sun}$] & $10^{-6}$ & \multicolumn{2}{c}{scaled to fit SED best} & 1 \\
                $R_{\text{in}}$~[au]                            & 40               & 5    & 60   & 2158                \\
                $R_{\text{out}}$~[au]                           & 200              & 100  & 500  & 1398                \\
                $\alpha$                                        & 1                & -1   & 3    & 80                  \\
                $s_{\text{min}}$~[\textmu m]                    & 3.73             & 0.1  & 10   & 403                 \\
                $s_{\text{max}}$~[\textmu m]                    & 1000             & 1000 & 1000 & 1                   \\
                $q$                                             & 3.5              & 2    & 5    & 150                 \\
                Inclination $i$~[$\degr$]                       & 0                & 0    & 0    & 1                   \\
                Porosity $\mathcal{P}$                          & 0.0 -- 0.9       & 0.0  & 0.0  & 1                   \\
                \hline
            \end{tabular}
        \end{table}
        
        \begin{table}
            \caption{Stellar properties of our sample. All values are taken from table~2 in \citet{pawellek-et-al-2014}.}
            \label{tab:luminosity}
            \centering
            \begin{tabular}{c c c}
                \hline\hline
                \rule{0pt}{2.5ex} $\nicefrac{L_{\star}}{\text{L}_{\sun}}$ & $\nicefrac{T_{\text{eff}}}{\text{K}}$ & $\nicefrac{M}{\text{M}_{\sun}}$ \\[1mm]
                \hline
                \rule{0pt}{2.5ex}
                0.012 &  3498  & 0.31 \\
                0.062 &  3600  & 0.48 \\
                0.41  &  5166  & 0.79 \\
                1.16  &  5930  & 1.04 \\
    %               1.25  &  5912  & 1.06 \\
                1.52  &  6155  & 1.12 \\
    %               1.83  &  6086  & 1.17 \\
    %               3.17  &  6490  & 1.35 \\
                3.44  &  6590  & 1.38 \\
    %               4.81  &  7380  & 1.51 \\
                4.87  &  6950  & 1.52 \\
    %               6.27  &  7070  & 1.62 \\
                7.04  &  7530  & 1.67 \\
    %               7.36  &  7200  & 1.69 \\
                10.3  &  7575  & 1.85 \\
    %               11.2  &  8325  & 1.89 \\
                11.7  &  8710  & 1.91 \\
    %               12.5  &  4815  & 1.94 \\
                13.2  &  8490  & 1.97 \\
    %               15.4  &  8550  & 2.05 \\
                15.5  &  8195  & 2.06 \\
    %               15.7  &  8400  & 2.07 \\
                16.0  &  9000  & 2.07 \\
    %               18.5  &  8925  & 2.16 \\
                24.9  &  10000 & 2.33 \\
    %               25.0  &  9350  & 2.33 \\
                26.0  &  9020  & 2.36 \\
    %               26.6  &  10190 & 2.37 \\
                31.3  &  9200  & 2.47 \\
    %               35.7  &  9770  & 2.56 \\
                51.8  &  9530  & 2.83 \\
    %               57.7  &  9220  & 2.91 \\
                58.2  &  9130  & 2.91 \\
    %               73.8  &  8010  & 3.10 \\
                \hline
            \end{tabular}
        \end{table}
    
    \subsection{Results}
    \label{sec:results}
        All results shown here are averaged over the five different fit runs of each disc; see end of \prettyref{sec:sand}. Furthermore, the results for the input data with $\mathcal{P} = 0.0 $ are used as a reference for all other results, as in this case the dust properties are identical for the simulated observations and the fit routine. Thus, in the following, we consider the ratio of the obtained parameter values of each run to the results of this special case. This has the advantage, that all variations from the input values occur as deviations from 1 and are completely due to the porosity and not to the fit process itself, as even with the same dust properties, the results of the fit can differ by a few percent from the input parameters because of the nature of the simulated annealing. The relative standard deviations of the reference case are less than 0.5 per cent for $q$, $\alpha$, $R_{\text{out}}$ and $M_{\mathrm{dust}}$, and less than 3.2 per cent for $s_{\text{min}}$ and $R_{\text{in}}$. These uncertainties are much less than the typical errors of best-fit models of real observational data due to measurement uncertainties and ambiguities. In the following, all uncertainties mentioned are absolute uncertainties of the normalized values.
        
\begin{figure*}
            \centering
            \resizebox{\hsize}{!}{\includegraphics{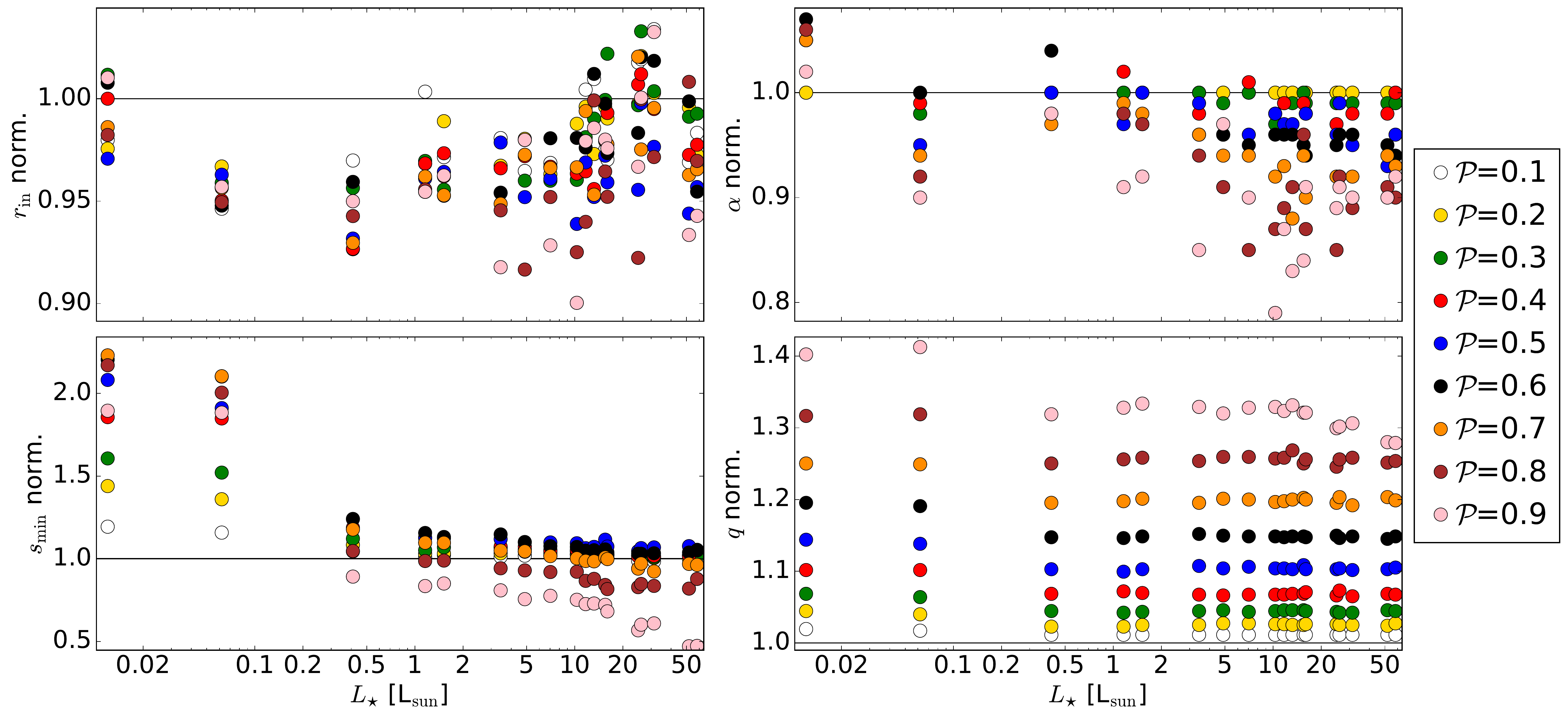}}
            \caption[caption]{Results of the fit processes for four parameters as functions of stellar luminosity, normalized to the reference case of $\mathcal{P}=0.0$.\\\hspace{\textwidth}
            \textit{Upper left:} Inner radius; \textit{Upper right:} Slope of radial density distribution; \textit{Lower left:} Minimum grain size; \textit{Lower right:} Slope of grain size distribution.
            Different colours indicate different grain porosities $\mathcal{P}$. The horizontal black line corresponds to a porosity of $\mathcal{P}=0.0$. For the sake of clarity, we omit to show error bars in this plot; see text for uncertainties.}
        \label{fig:results}
        \end{figure*}

        In \prettyref{fig:results}, we show the normalized (to the compact case) mean value of the fit results as a function of stellar luminosity. Thus, values smaller than one correspond to an underestimation of the parameter and a value larger than one to an overestimation compared to non-porous grains. In the following, we will discuss the results for every parameter, separately.\vspace*{5mm}
        
        \noindent\textit{Inner radius $R_{in}$} (\prettyref{fig:results}, upper left): The inner radius shows no obvious trend with luminosity, but a minor trend with porosity. A more porous composition leads to a smaller inner radius, down to 90 per cent of the original value. However, the scatter is very large, as well as the uncertainties. The inner rim is not well determined with uncertainties ranging between 1 per cent and 8 per cent constantly for all porosities. This signifies, that this trend is unreliable, as the uncertainties are in the order of the maximum deviation.
        \vspace*{5mm}
        
        \noindent\textit{Radial density profile (exponent $\alpha$)} (\prettyref{fig:results}, upper right): Although the uncertainties for $\mathcal{P} \ge 0.3$ are as high as for the inner radius, the trend with porosity is much more obvious. Moreover, for smaller porosities $\alpha$ is very well determined, i.e. all five fit runs for each star result in the same value and hence no statistical uncertainty can be derived.
        \vspace*{5mm}
        
\begin{figure}
            \centering
            \resizebox{\hsize}{!}{\includegraphics{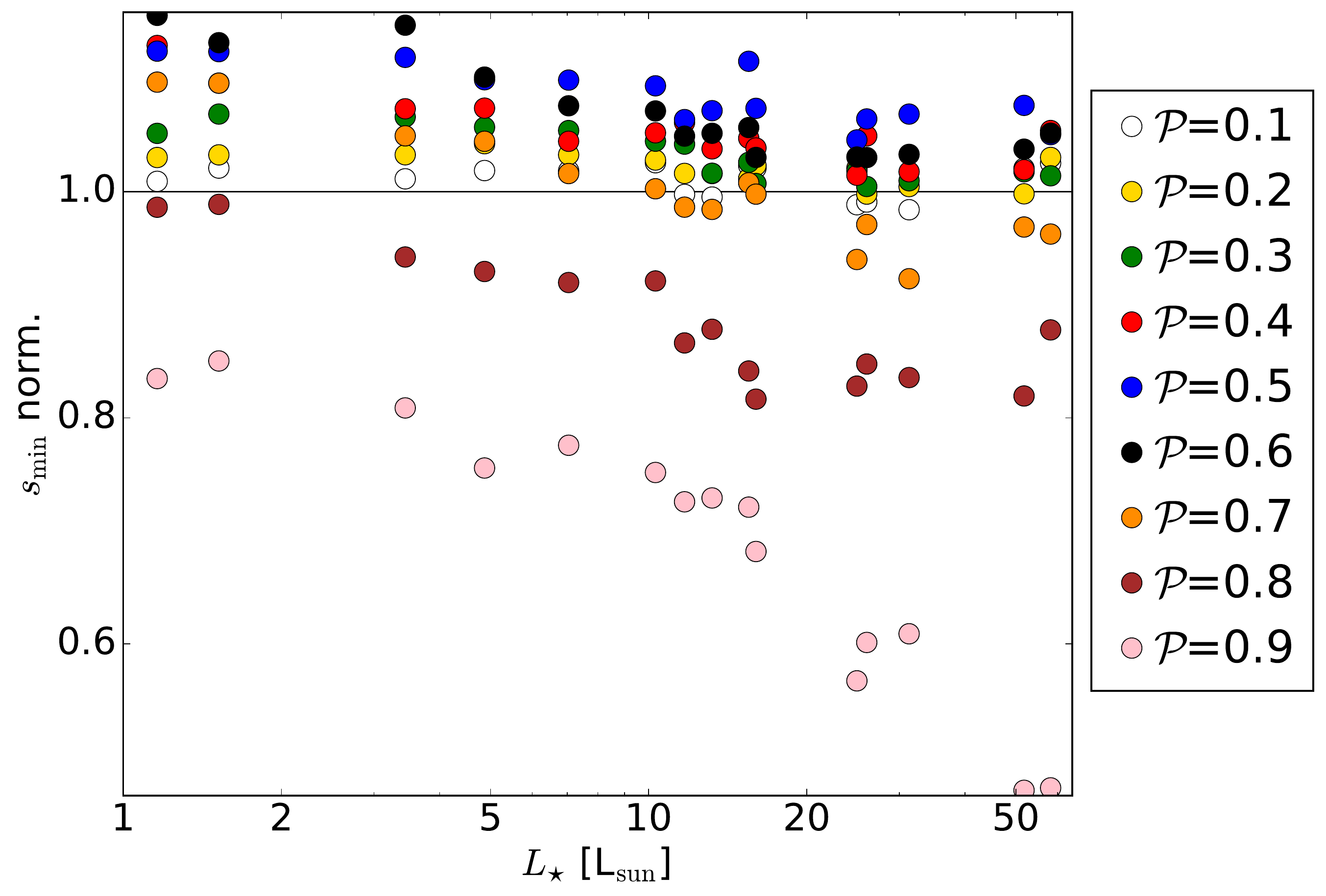}}
            \caption[caption]{Zoomed-in version of lower left panel in \prettyref{fig:results}.\\\hspace{\textwidth}
            Different colours indicate different grain porosities $\mathcal{P}$. The horizontal black line corresponds to a porosity of $\mathcal{P}=0.0$. For the sake of clarity, we omit to show error bars in this plot; see text for uncertainties.}
        \label{fig:results_smin}
        \end{figure}

        \noindent\textit{Minimum grain size $s_{min}$} (\prettyref{fig:results}, lower left): This plot clearly shows a dependency on both the stellar luminosity and the dust grain porosity. At first, the deviations increase with increasing porosity but then decrease again for $\mathcal{P}\gtrsim0.5$. For larger luminosities, the normalized $s_{min}$ decreases for all porosities. The minimum grain size is overestimated by a factor of more than two for stars WITH $L_{\star} \la 1~\text{L}_{\sun}$. \prettyref{fig:results_smin} is a zoomed-in version of this plot for the brighter stars. The luminosity dependency is still visible and best seen for large porosities. The uncertainties are less than 5 per cent in most cases for stellar luminosities of $\la 1~\text{L}_{\sun}$. However, for the dimmest star in our sample the uncertainties reach values of up to 15 per cent.
        \vspace*{5mm}
        
        \noindent\textit{Slope of grain size distribution $q$} (\prettyref{fig:results}, lower right): Here, the porosity has the most obvious effect on the fit results. For all luminosities and porosities, the size distribution slope is overestimated, i.e. larger than the correct value of 3.5. The slope is increasing with increasing porosity, up to more than 40 per cent above the original value, but no trend can be seen for stellar luminosity, although $q$ is slightly higher for the two dimmest stars with $L_{\star} < 0.1~\text{L}_{\sun}$. The uncertainties are less than 1 per cent almost all runs.
        \vspace*{5mm}
        
\begin{figure}
            \centering
            \resizebox{\hsize}{!}{\includegraphics{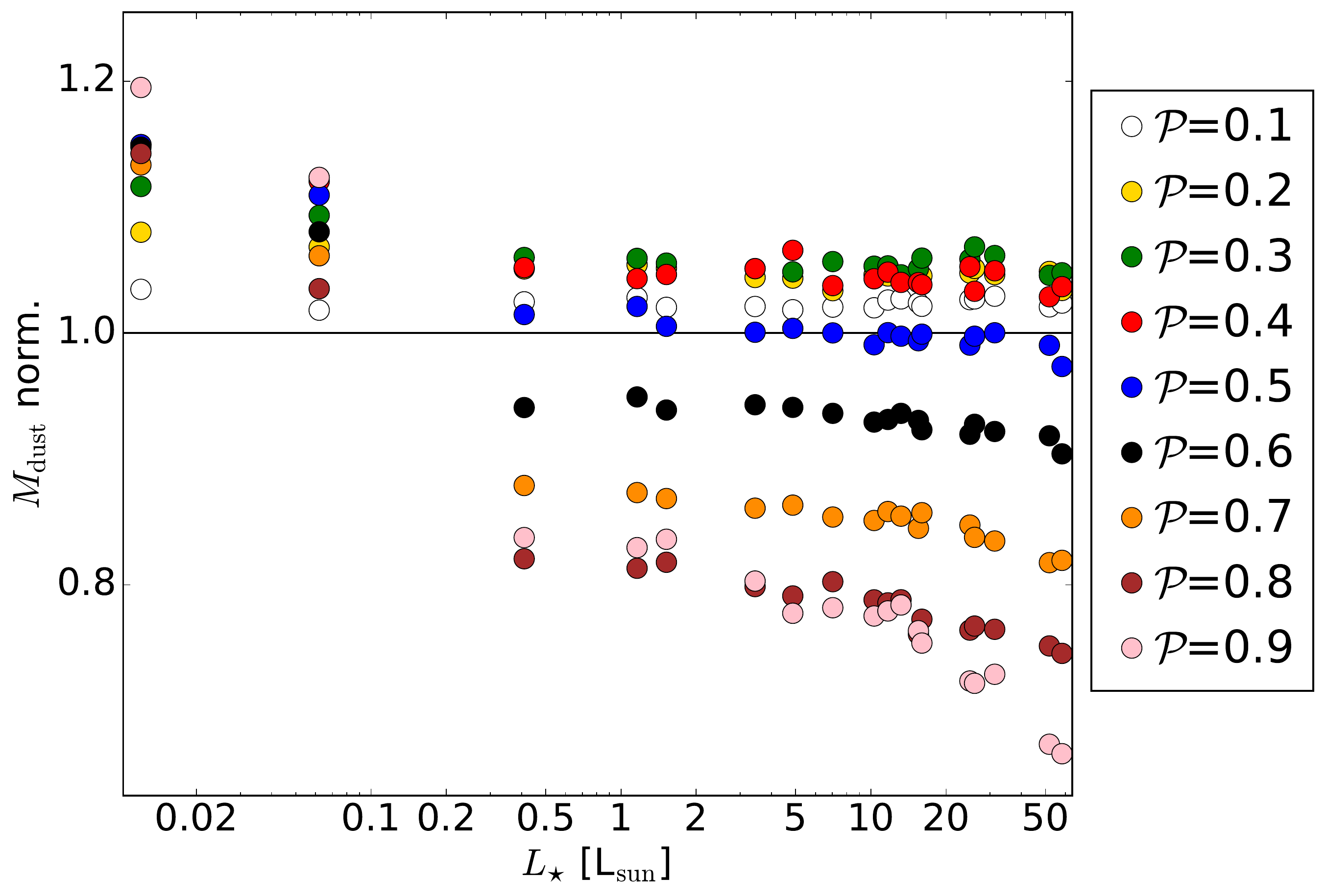}}
            \caption[caption]{Results of the fit process for the dust mass as function of stellar luminosity, normalized to the reference case of $\mathcal{P}=0.0$.\\\hspace{\textwidth}
            Different colours indicate different grain porosities $\mathcal{P}$. The horizontal black line corresponds to a porosity of $\mathcal{P}=0.0$. For the sake of clarity, we omit to show error bars in this plot; see text for uncertainties.}
        \label{fig:results_mass}
        \end{figure}

        \noindent\textit{Dust mass $M_{\mathrm{dust}}$} (\prettyref{fig:results_mass}): The deviations from the input value increase for increasing porosities, up to an overestimation of 20 per cent and an underestimation of more than 30 per cent for $\mathcal{P}=0.9$. The porosity dependence is clearly visible. The mass was overestimated for dim stars and underestimated for stars with $L_{\star} \ga 1~\text{L}_{\sun}$. In all cases, the uncertainties are very low, in the range of 1 -- 2 per cent.
        \vspace*{5mm}
        
        \noindent\textit{Outer radius $R_{out}$}: Because of the small FWHM of the Gaussian and the large outer radius, the latter can be determined with high accuracy from the radial profiles. Thus, nearly all runs result in the correct input value, i.e. all data points are equal to one or differ by less than 1 per cent with uncertainties of less than 0.7 per cent. Because of this, we omit a corresponding plot, as this would give no further information.\vspace*{5mm}
        
        The spatial dust distribution is well-determined, which is due to the high resolution of our re-emission images, as mentioned above. This has the advantage, that existent ambiguities, such as the connection between inner radius and minimum grain size, are resolved, and trends are due to porosity only. However, this spatial resolution is not yet achieved for most of the known debris discs, especially for those that are farther away. Therefore, these results are only valid for well-resolved discs.
        
\section{Conclusion}
\label{sec:conclusion}
    We investigated the influence of porosity on the theoretical blowout size and on the analysis of debris disc observations. We were able to derive an analytical expression for the blowout size $s_{\text{blow}}$ as a function of stellar luminosity and porosity for grains composed of astrosil. We found that the commonly used approximation results in a systematically underestimated blowout size and a slope that is too steep compared to a more detailed description in the case of compact particles. Thus, this approximation introduced by \citet{burns-et-al-1979} should be discarded and replaced with the equaly simple, but more accurate \prettyref{eq:sblow}. This equation is of course only valid for astronomical silicate, spherical particles and porosities that are low enough ($\mathcal{P}\lesssim0.6$). Furthermore, we find that for high porosities the blowout size increases by a factor of 5 for high luminosities ($L_{\star} \approx 50~\text{L}_{\sun}$) and by a factor of 1.4 for low stellar luminosities ($L_{\star} \approx 1~\text{L}_{\sun}$) and medium porosities ($\mathcal{P}=0.4$). The higher the porosities, the lower the $\beta$-ratio. Thus, small particles can survive even around luminous stars, where compact particles with the same size would be expelled by the radiation pressure.
    
    In the second part, we showed that the analysis of the geometrical structure of a debris disc is barely affected by porous grains. As this information is mostly derived from the well-resolved radial brightness profiles, this result fulfills the expectations, especially for the outer radius. However, the grain size distribution and the minimum grain size show a dependency on porosity. A fit process that does not consider fluffy grains overestimates the slope $q$ of the grain size distribution by up to 40 per cent and $s_{\text{min}}$ by up to 230 per cent for very dim stars and underestimates $s_{\text{min}}$ by a factor of up to 0.4 for luminosities $L_{\star}\ge1\text{L}_{\sun}$ and porosities $\mathcal{P}\approx0.9$. Thus, the actual minimum grain size can be smaller or larger than the value derived from the analysis, depending on luminosity and porosity. Besides, the extent of these deviations can exceed the typical best-fit uncertainties of debris disc modelling \citep{loehne-et-al-2012}.
    
    \citet{pawellek-et-al-2014} stated that the ratio of minimum grain size to blowout size is very high ($\approx10$) for solar-type stars and decreases with increasing luminosity, reaching unity for $L_{\star}\approx 50~\text{L}_{\sun}$. In this study, we showed that neglecting porosity will change both quantities in such a way that this ratio can be significantly decreased by a factor of 2 -- 4. Thus, the influence of porosity is not large enough to explain why the minimum grain size is so much higher than expected, although it moderates the trend. Thus, other effects with influence on the ratio, such as the imbalance of dust destruction and production rates \citep{thebault-wu-2008}, different compositions \citep{pawellek-krivov-2015}, and the surface energy constraint during collisional dust production \citep{pawellek-krivov-2015,thebault-2016}, need to be considered, too.
    
\section*{Acknowledgement}
    
    This research was funded through the DFG grants WO 857/13-1 and WO 857/15-1.

%%%%%%%%%%%%%%%%%%%%%%%%%%%%%%%%%%%%%%%%%%%%%%%%%%

%%%%%%%%%%%%%%%%%%%% REFERENCES %%%%%%%%%%%%%%%%%%

\bibliographystyle{../../mnras/mnras}
\bibliography{./_lit}

% Don't change these lines
\bsp    % typesetting comment
\label{lastpage}
\end{document}